# A PROPOSAL TO INTEGRATE DATA ENVELOPMENT ANALYSIS AND LE CHATELIER PRINCIPLE

*Filippo Elba[1], research and teaching assistant at Dep. Economics & Management – University of Florence, Florence, Italy.*

**Abstract**
*The article aims to estimate the short and long run efficient production frontiers for firm operating in an industry. The Le Chatelier principle represents the theoretical framework, while the tool used to find the frontiers is the non-parametric Data Envelopment Analysis technique. The proposal tries to overcome the main limitations that other efforts done in this context seem to have.*

## 1. Introduction

Le Chatelier principle links short and long period production functions for firms operating in an industry. Starting from previous works, a method based on the non-parametric technique Data Envelopment Analysis (DEA) is here presented in order to find the two functions.

The present work is so organized: in section 2 Le Chatelier principle is described in details; in section 3 there is a brief description of the DEA technique and how it has already been adopted in the past in this context; in section 4 a description of our proposal is done; conclusions follow.

## 2. The Le Chatelier principle

The Le Chatelier principle applied to economics (Samuelson, 1954) implies that firms' short and long run production functions are different. The difference is due to the fact that some production factors are fixed in the short period and variable in the long one.

For simplicity, we consider two production factors only, i.e. labour (L) and capital (K): in the long term, firms can manage both factors; otherwise, in the short term, firms can manage labour factor only, being done the capital level (e.g., in the short run is extremely difficult changing plants or machineries number). In consequence of what happens in the short run, firms are subject to the so-called law of diminishing marginal returns. Formally, the short term production function is:

$$Q_{BP} = F(L, \overline{K}) \qquad (1)$$

$Q_{BP}$ represents the short period production level; $\overline{K}$ represents the fixed capital level; L the labour quantity.

The reasons explaining how firms choose the short term capital quantity are described by Johansen (1972). His work shows the relationships between the individual entrepreneurial choices in short and long run and the individual and aggregated firms supplies. With regards to the first couple, the scholar puts in evidence that short term behaviour is determined by "crystallisation" of strategic choices made before that the company has been created. After that moment, the capital quantity is fixed and the labour quantity is variable. The factors and final goods' prices determine the optimal K and L quantities so as to produce the Q level that maximizes profits.

Whether the capital crystallisation belongs to short perspective, this is not true in the long one: the prices neutrality makes labour and capital both variable, then the (1) becomes:

$$Q_{LP} = F(L, K) \qquad (2)$$

In the long run, the firm can benefit from returns to scale, which depend from capital level.

What has been said determines the classical graphical representation of the average total cost in the short and long periods. Whether in the short run the objective of costs minimization depends from capital quantity (1) and then there are many local minimums (dotted lines in Fig. 1), in the long run a minimum value of average cost per unit exists and it is the same one for all firms operating in an industry (continuous line in Fig. 1).

---
[1] Corresponding author. E-mail: filippo.elba@unifi.it.

**Fig. 1: Short and long run curves of costs per unit produced**

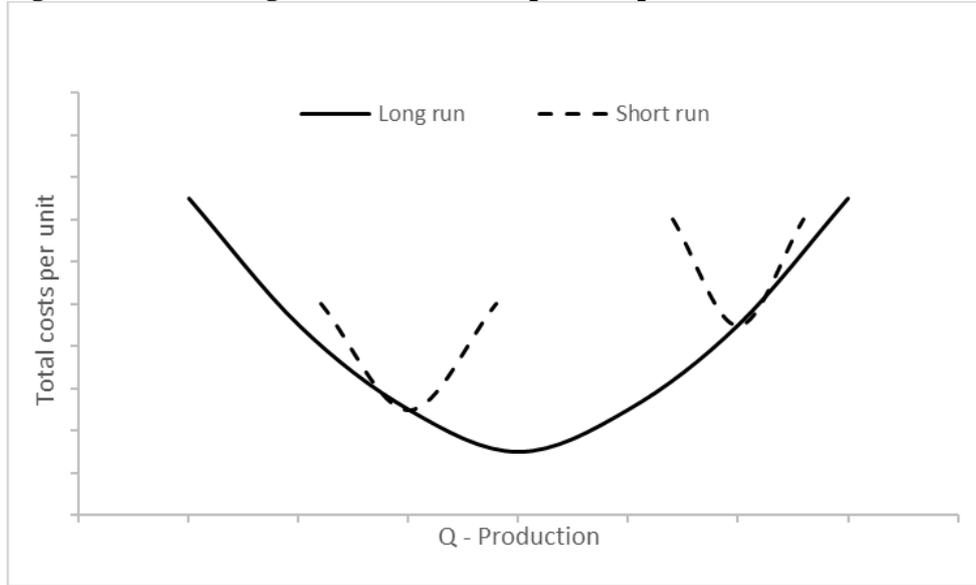

## 3. DEA and Le Chatelier principle

DEA is a non-parametric technique, based on linear programming, useful to find the efficient production function. Considering that x represents the values of m inputs of n production units and y the values of their s outputs, Charnes et al. (1978) formalize the dual version of the optimization problem, called DEA-CCR, as follows[2]:

$$\text{Min } \theta - \left( \sum_{i=1}^{m} s_i^- + \sum_{r=1}^{s} s_r^+ \right) \quad (3)$$

s.t.

$$\sum_{j=1}^{n} x_{ij} \lambda_j + s_i^- = \theta x_{io}$$

$$\sum_{j=1}^{n} y_{rj} \lambda_j - s_r^+ = y_{ro}$$

$$\lambda_j > 0$$

$\theta$ represents radial inefficiency (i.e. indistinctly referred to all inputs) given outputs level. It can assume values between 0 and 1. $s_i^-$ and $s_r^+$ represent slacks, such as further inefficiencies referred to singular input and/or output. For this reason, they can be negative or positive: for inputs, slacks indicate a further reduction, in addition to the radial one; for outputs indicate an increase.

The (3) implies constant returns to scale. Introducing the condition $\sum_{j=1}^{n} \lambda_j = 1$, the model can take into account variable returns to scale (Banker at al., 1984).

On this regards, a clarification is necessary with respect to "fixity" or "variability" nature of inputs in DEA. When all inputs are variable, we can estimate: efficient production frontier with constant returns to scale (1); with variable returns to scale (adding the constrain on lambdas' summation); calculating the scale order through ratio between the former on the latter. Then we can read DEA results with constant returns to scale as reachable in a long term, because reaching them implies to modify scale order, and results with variable returns to scale as reachable in a short term, because they do not need a scale change.

---

[2] It is here presented the "input minimizing" DEA version. The choice is motivated by the fact that capital is an input in production processes. For information about "output maximizing" version and its primal formulation: Charnes *et al.* (1978).

What has been said cannot be considered acceptable when capital is taken into account as input: how could we consider an inefficiency with variable returns to scale condition knowing that capital is fixed in the short run? Even if it was possible to change its value, how could we read results knowing that whether we improved capital quantity, scale order would change (which is the same assumption beyond the constant returns to scale evaluation)? Nevertheless, capital is often considered as input in DEA models with variable returns to scale and frontiers so obtained are often interpreted as the short run efficient ones (e.g.: Hawdon, 2003).

Scholars have already dealt with the problem of evaluating efficient production frontier using DEA under the Le Chatelier principle assumptions. In some cases, they have explicitly referred to the principle (Chen & Huang, 2009; Diewert & Mendoza; 2007), but more often they have generally referred to differences between short and long run production functions considering capital inputs as "quasi-fixed" in the first case (e.g.: Prior & Filimon 2002)[3].

The Chen & Huang (2009)'s proposal seems to be one among the best ones. For that concerning the efficient frontier estimate in a long ran perspective, the scholars consider the DEA-CCR version the most appropriate(3). As already said, in the long period all kind of input variables can be managed, so there are no differences between capital and labour. So as to determine the short run productivity frontier, the scholars adopt a DEA-CCR model considering capital as non-discretionary(ND) input. In this way, they leave out the possibility of intervention on it (by the firm's management):

$$\text{Min } \theta - \left( \sum_{i \in D} s_i^- + \sum_{r=1}^{s} s_r^+ \right) \quad (4)$$

s.t.

$$\sum_{j=1}^{n} x_{ij} \lambda_j + s_i^- = \theta x_{io} \; ; i \in D$$

$$\sum_{j=1}^{n} x_{ij} \lambda_j + s_i^- = x_{io} \; ; i \in ND$$

$$\sum_{j=1}^{n} y_{rj} \lambda_j - s_r^+ = y_{ro} \; ; \quad \lambda_j > 0$$

As consequence, in the short run, firms are compared with them using at least the same quantity of capital. Then the capital quantity seems to have the same role that in the variable returns to scale DEA has the constrain on lambdas' summation.

This solution seems to have some limitations: since capital is dealt as non-discretionary input, de facto it is not considered in the object function (4) that represents the efficient production frontier, in contrast with the (1). Then, Chen & Huang, not considering capital in productivity frontier, adopt a "classical" solution without capital input, while in long run they consider the variable. Not only it seems to be contradictory, but this solution do not permit to evaluate differences between the two models results (for further details on the relationship between DEA scores and number of problem dimensions – number of inputs and outputs in a model – see Lopez et al., 2016).

---

[3] The cited works deal not only with the matter of finding the two frontiers, but also with that to calculate the allocative efficiency, a topic that this article does not face.

## 4. A model based on Additive DEA CRS for short run

The proposal consists in solving the problem of estimating the short run productivity frontier with an iterative model based on Additive DEA CRS[4].

The iterative process includes the following steps:
1. realisation of a model based on Additive DEA CRS with capital among inputs and generation of results;
2. for each firm, checking of values of $s_i^-$ referred to capital input:
    I. whether it is equal to 0, jump to point 3;
    II. whether it is different from 0: find the firm's peers; leave out them from the units sample; come back to point 1;
3. since the firm is at least "weak efficient" (i.e., it has the characteristics in point 2.I), its short term inefficiencies are correctable improving variables presenting slacks (with the exception of capital slack that has to be 0).

In this way a short run efficient productivity frontier is given for each firm and capital appears as input in the objective function (for further details, see the example in Appendix).

Adopting a slack-based DEA perspective (Tone, 2001), a comparison between the short and long term models results is possible.

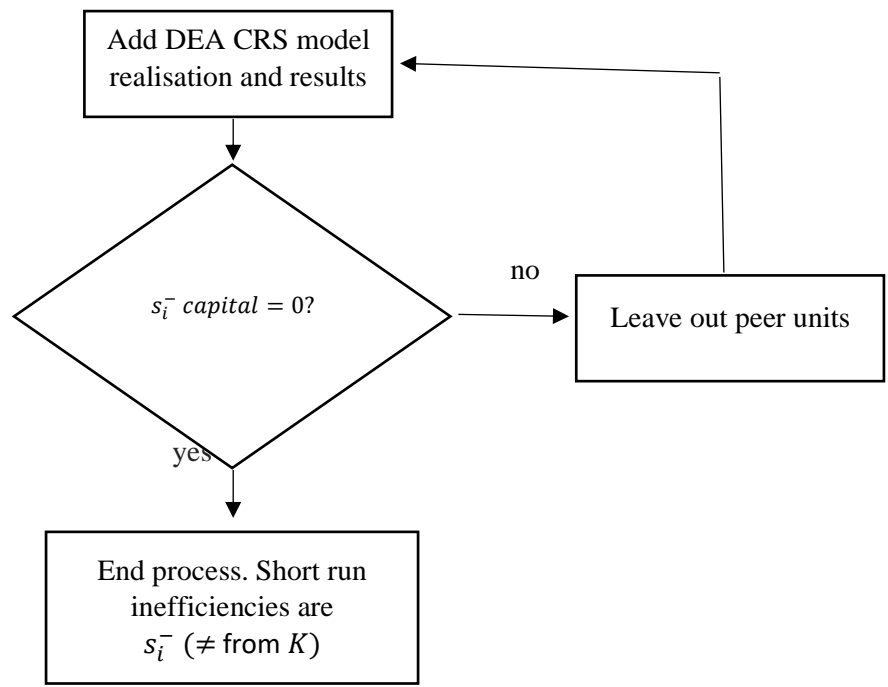

**Fig. 2: Iterative process based on Additive DEA CRS to estimate short run productivity frontier**

## 5. Conclusions

The purpose of the article is presenting a practical solution to the problem of estimating short and long run efficient production frontiers, in the Le Chatelier's sense, using DEA. Starting from the solution proposed by Chen & Huang (2009), a model is here presented in order to overcome the main limitations of previous efforts. For future works, we aim to answer the question of how to use the model in a dynamic framework so as to distinguish gains due to the scale improvements and the ones due to technological progress.

---

[4] The additive model is similar to that presented in paragraph 3, with the difference that only slacks are considered, not distinguishing the radial component θ and its minimization.

## Appendix

Data example generation:

```
A<-c(10,20,30)
B<-c(10,12,30)
C<-c(12,13,30)
D<-c(10,15,30)
E<-c(20,30,30)
G<-c(20,40,30)
H<-c(18,10,30)
I<-c(19,10,30)
J<-c(11,11,29)

data<-as.matrix(rbind(A,B,C,D,E,G,H,I,J))
```

Additive DEA CRS model on the basis of Bogetoft & Otto (2018) R package:

```
modCrs<-dea.add(X=data[,1:2],Y=data[,3],RTS="crs")
results<-cbind(modCrs$lambda,modCrs$sx,modCrs$sy)
results
```

Iterative process as described above. The first dataset column represents the capital input:

```
iterativeRes<-list()
df1<-data
res<-results
getworse<-c(max(df1[,1]),max(df1[,2]),min(df1[,3])) # to get worse the situation of the peers excluding them from the frontier
for(i in 1:nrow(data)){
  res<-results
  df1<-data
  while(round(res[i,nrow(res)+1],3)>0){
    peers<-which(res[i,1:nrow(res)]>0,arr.ind = T)
    names(peers)<-NULL
    p<-c(peers)
    for(ii in p){
      df1[ii,]<-getworse
      }
    mod<-dea.add(X=as.matrix(df1[,1:2]),
                 Y=as.matrix(df1[,3]), RTS="crs")
    risu1<-cbind(mod$lambda,mod$sx,mod$sy)
    }
  iterativeRes[[i]]<-res
    }
iterativeRes
```

Here the results for the first model (that is the long run efficiency):

```
      L1       L2 L3 L4 L5 L6 L7 L8 L9      sx1 sx2  sy1
[1,]   0 1.000000  0  0  0  0  0  0  0 0.000000    8  0.0
[2,]   0 1.000000  0  0  0  0  0  0  0 0.000000    0  0.0
[3,]   0 1.083333  0  0  0  0  0  0  0 1.166667    0  2.5
[4,]   0 1.000000  0  0  0  0  0  0  0 0.000000    3  0.0
[5,]   0 2.000000  0  0  0  0  0  0  0 0.000000    6 30.0
[6,]   0 2.000000  0  0  0  0  0  0  0 0.000000   16 30.0
[7,]   0 0.000000  0  0  0  0  1  0  0 0.000000    0  0.0
[8,]   0 0.000000  0  0  0  0  1  0  0 1.000000    0  0.0
[9,]   0 0.000000  0  0  0  0  0  0  1 0.000000    0  0.0
```

Here the short run efficiency, which differs from the above especially for the units that in the long run efficiency presents sx1>0 (that is the slack on the capital input):

```
[[1]]
      L1       L2 L3 L4 L5 L6 L7 L8 L9      sx1 sx2  sy1
[1,]   0 1.000000  0  0  0  0  0  0  0 0.000000    8  0.0
[2,]   0 1.000000  0  0  0  0  0  0  0 0.000000    0  0.0
[3,]   0 1.083333  0  0  0  0  0  0  0 1.166667    0  2.5
[4,]   0 1.000000  0  0  0  0  0  0  0 0.000000    3  0.0
[5,]   0 2.000000  0  0  0  0  0  0  0 0.000000    6 30.0
[6,]   0 2.000000  0  0  0  0  0  0  0 0.000000   16 30.0
[7,]   0 0.000000  0  0  0  0  1  0  0 0.000000    0  0.0
[8,]   0 0.000000  0  0  0  0  1  0  0 1.000000    0  0.0
[9,]   0 0.000000  0  0  0  0  0  0  1 0.000000    0  0.0

[[2]]
      L1       L2 L3 L4 L5 L6 L7 L8 L9      sx1 sx2  sy1
[1,]   0 1.000000  0  0  0  0  0  0  0 0.000000    8  0.0
[2,]   0 1.000000  0  0  0  0  0  0  0 0.000000    0  0.0
[3,]   0 1.083333  0  0  0  0  0  0  0 1.166667    0  2.5
[4,]   0 1.000000  0  0  0  0  0  0  0 0.000000    3  0.0
[5,]   0 2.000000  0  0  0  0  0  0  0 0.000000    6 30.0
[6,]   0 2.000000  0  0  0  0  0  0  0 0.000000   16 30.0
[7,]   0 0.000000  0  0  0  0  1  0  0 0.000000    0  0.0
[8,]   0 0.000000  0  0  0  0  1  0  0 1.000000    0  0.0
[9,]   0 0.000000  0  0  0  0  0  0  1 0.000000    0  0.0
```

```
[[3]]
     L1 L2 L3 L4 L5 L6 L7 L8       L9 sx1 sx2       sy1
[1,]  0  0  0  1  0  0  0  0 0.000000   0   5  0.000000
[2,]  0  0  0  0  0  0  0  0 1.818182   0  20 23.727273
[3,]  0  0  0  0  0  0  0  0 1.090909   0   1  1.636364
[4,]  0  0  0  1  0  0  0  0 0.000000   0   0  0.000000
[5,]  0  0  0  0  0  0  0  0 1.818182   0  10 22.727273
[6,]  0  0  0  0  0  0  0  0 1.818182   0  20 22.727273
[7,]  0  0  0  0  0  0  1  0 0.000000   0   0  0.000000
[8,]  0  0  0  0  0  0  1  0 0.000000   1   0  0.000000
[9,]  0  0  0  0  0  0  0  0 1.000000   0   0  0.000000

[[4]]
     L1       L2 L3 L4 L5 L6 L7 L8 L9      sx1 sx2  sy1
[1,]  0 1.000000  0  0  0  0  0  0  0 0.000000   8  0.0
[2,]  0 1.000000  0  0  0  0  0  0  0 0.000000   0  0.0
[3,]  0 1.083333  0  0  0  0  0  0  0 1.166667   0  2.5
[4,]  0 1.000000  0  0  0  0  0  0  0 0.000000   3  0.0
[5,]  0 2.000000  0  0  0  0  0  0  0 0.000000   6 30.0
[6,]  0 2.000000  0  0  0  0  0  0  0 0.000000  16 30.0
[7,]  0 0.000000  0  0  0  0  1  0  0 0.000000   0  0.0
[8,]  0 0.000000  0  0  0  0  1  0  0 1.000000   0  0.0
[9,]  0 0.000000  0  0  0  0  0  0  1 0.000000   0  0.0

[[5]]
     L1       L2 L3 L4 L5 L6 L7 L8 L9      sx1 sx2  sy1
[1,]  0 1.000000  0  0  0  0  0  0  0 0.000000   8  0.0
[2,]  0 1.000000  0  0  0  0  0  0  0 0.000000   0  0.0
[3,]  0 1.083333  0  0  0  0  0  0  0 1.166667   0  2.5
[4,]  0 1.000000  0  0  0  0  0  0  0 0.000000   3  0.0
[5,]  0 2.000000  0  0  0  0  0  0  0 0.000000   6 30.0
[6,]  0 2.000000  0  0  0  0  0  0  0 0.000000  16 30.0
[7,]  0 0.000000  0  0  0  0  1  0  0 0.000000   0  0.0
[8,]  0 0.000000  0  0  0  0  1  0  0 1.000000   0  0.0
[9,]  0 0.000000  0  0  0  0  0  0  1 0.000000   0  0.0

[[6]]
     L1       L2 L3 L4 L5 L6 L7 L8 L9      sx1 sx2  sy1
[1,]  0 1.000000  0  0  0  0  0  0  0 0.000000   8  0.0
[2,]  0 1.000000  0  0  0  0  0  0  0 0.000000   0  0.0
[3,]  0 1.083333  0  0  0  0  0  0  0 1.166667   0  2.5
[4,]  0 1.000000  0  0  0  0  0  0  0 0.000000   3  0.0
[5,]  0 2.000000  0  0  0  0  0  0  0 0.000000   6 30.0
[6,]  0 2.000000  0  0  0  0  0  0  0 0.000000  16 30.0
[7,]  0 0.000000  0  0  0  0  1  0  0 0.000000   0  0.0
[8,]  0 0.000000  0  0  0  0  1  0  0 1.000000   0  0.0
[9,]  0 0.000000  0  0  0  0  0  0  1 0.000000   0  0.0

[[7]]
     L1       L2 L3 L4 L5 L6 L7 L8 L9      sx1 sx2  sy1
[1,]  0 1.000000  0  0  0  0  0  0  0 0.000000   8  0.0
[2,]  0 1.000000  0  0  0  0  0  0  0 0.000000   0  0.0
[3,]  0 1.083333  0  0  0  0  0  0  0 1.166667   0  2.5
[4,]  0 1.000000  0  0  0  0  0  0  0 0.000000   3  0.0
[5,]  0 2.000000  0  0  0  0  0  0  0 0.000000   6 30.0
[6,]  0 2.000000  0  0  0  0  0  0  0 0.000000  16 30.0
[7,]  0 0.000000  0  0  0  0  1  0  0 0.000000   0  0.0
[8,]  0 0.000000  0  0  0  0  1  0  0 1.000000   0  0.0
[9,]  0 0.000000  0  0  0  0  0  0  1 0.000000   0  0.0

[[8]]
     L1       L2 L3 L4 L5 L6 L7 L8 L9      sx1 sx2  sy1
[1,]  0 1.000000  0  0  0  0  0  0  0 0.000000   8  0.0
[2,]  0 1.000000  0  0  0  0  0  0  0 0.000000   0  0.0
[3,]  0 1.083333  0  0  0  0  0  0  0 1.166667   0  2.5
[4,]  0 1.000000  0  0  0  0  0  0  0 0.000000   3  0.0
[5,]  0 2.000000  0  0  0  0  0  0  0 0.000000   6 30.0
[6,]  0 2.000000  0  0  0  0  0  0  0 0.000000  16 30.0
[7,]  0 2.000000  0  0  0  0  0  0  0 0.000000  16 31.0
[8,]  0 0.000000  0  0  0  0  1  0  0 0.000000   0  0.0
[9,]  0 0.000000  0  0  0  0  0  0  1 0.000000   0  0.0

[[9]]
     L1       L2 L3 L4 L5 L6 L7 L8 L9      sx1 sx2  sy1
[1,]  0 1.000000  0  0  0  0  0  0  0 0.000000   8  0.0
[2,]  0 1.000000  0  0  0  0  0  0  0 0.000000   0  0.0
[3,]  0 1.083333  0  0  0  0  0  0  0 1.166667   0  2.5
[4,]  0 1.000000  0  0  0  0  0  0  0 0.000000   3  0.0
[5,]  0 2.000000  0  0  0  0  0  0  0 0.000000   6 30.0
[6,]  0 2.000000  0  0  0  0  0  0  0 0.000000  16 30.0
[7,]  0 0.000000  0  0  0  0  1  0  0 0.000000   0  0.0
[8,]  0 0.000000  0  0  0  0  1  0  0 1.000000   0  0.0
[9,]  0 0.000000  0  0  0  0  0  0  1 0.000000   0  0.0
```